\documentclass{emulateapj}

\newcommand{\mdot}{M$_{\odot}$ yr$^{-1}$}
\newcommand{\ldot}{L$_{\odot}$}
\newcommand{\um}{$\mu$m~}
\newcommand{\ums}{$\mu$m}

\slugcomment{ApJSup accepted (Spitzer Special Issue)}

\shorttitle{Faint Galaxy Spectra}
\shortauthors{Higdon et al.}

\begin{document}

\title{First mid-infrared spectrum of a faint high-z galaxy: Observations of
       CFRS 14.1157 with the Infrared Spectrograph\footnotemark [1] on
       the Spitzer Space Telescope \footnotemark [2]}

\author{S. J. U. Higdon\altaffilmark{3}, D. Weedman\altaffilmark{3}, 
J. L. Higdon\altaffilmark{3}, T. Herter\altaffilmark{3}, 
V. Charmandaris\altaffilmark{3,4}, J. R. Houck\altaffilmark{3}, 
 B. .T. Soifer\altaffilmark{5}, B. R. Brandl\altaffilmark{6}, 
L. Armus\altaffilmark{5} \& L. Hao\altaffilmark{3}}

\footnotetext [1] {The IRS was a
       collaborative venture between Cornell University and Ball
       Aerospace Corporation funded by NASA through the Jet Propulsion
       Laboratory and the Ames Research Center.}
\footnotetext [2] {Based on
       observations obtained with the Spitzer Space Telescope, which
       is operated by JPL, California Institute of Technology for the
       National Aeronautics and Space Administration.}

\altaffiltext{3}{Astronomy Department, Cornell University, Ithaca, NY 14853}
\altaffiltext{4}{Chercheur Associ\'e, Observatoire de Paris, F-75014, Paris, France}
\altaffiltext{5}{Spitzer Science Center, California Institute of Technology, 220-6, Pasadena, CA 91125}
\altaffiltext{6}{Leiden University, 2300 RA Leiden, The Netherlands}

\begin{abstract}

The unprecedented sensitivity of the Infrared Spectrograph on the
Spitzer Space Telescope allows for the first time the measurement of
mid-infrared spectra from 14 \um to 38 \um of faint high-z galaxies.
This unique capability is demonstrated with observations of sources
having 16 \um fluxes of 3.6 mJy (CFRS 14.1157) and 0.35 mJy (CFRS
14.9025). A spectral-fitting technique is illustrated which determines
the redshift by fitting emission and absorption features
characteristic of nearby galaxies to the spectrum of an unknown
source. For CFRS 14.1157, the measured redshift is z = 1.00 $\pm$ 0.20
in agreement with the published result of z = 1.15. The spectrum is
dominated by emission from an AGN, similar to the nucleus of NGC 1068,
rather than a typical starburst with strong PAH emission like
M82. Such spectra will be crucial in characterizing the nature of
newly discovered distant galaxies, which are too faint for optical
follow-up.

\end{abstract}


\section{Introduction}

Observational studies of the star formation history of the universe
indicate that the star formation density rises rapidly from the
present to z $\sim$ 1 with very uncertain behavior at higher redshifts
(e.g., Elbaz \& Cesarsky 2003 and references therein).  Determining
the luminosity functions and redshift distribution of dusty starburst
galaxies is crucial to deducing the star formation history of the
universe. The infrared and submillimeter windows are ideal hunting
grounds for discovering these sources as most of the luminosity from
starbursts is re-radiated by dust at infrared and submillimeter
wavelengths.  A closely related problem is discriminating between
starbursts and AGNs as the fundamental power source for heating this
dust.

The Infrared Space Observatory (ISO) detected such infrared galaxies
down to $\sim$ 0.1 mJy at $\sim$15 $\mu$m within some small survey
areas, such as the Canada-France Redshift Survey (CFRS) fields number
1415+52 and 0300+00 (Flores et al. 1999) and the Hubble Deep Field
North (HDFN) \citep{aussel99}. The redshift distribution of the ISO
sources peak at z $\sim$ 1.  These results are used to infer the
evolution of star formation under the assumption that these galaxies
are primarily powered by starbursts.  In some cases, this assumption
can be confirmed by optical emission line ratios, but in many cases -
especially the most luminous sources - the optical spectra show
indicators of AGN \citep{veil99}, so it cannot be determined what
fraction of the infrared luminosity derives from a starburst compared
to the fraction from the AGN.

If z $>$ 1 galaxies are similar to those in the local universe we
would expect starbursts to have strong polycyclic-aromatic hydrocarbon
features (PAH, at rest wavelengths 6.2 \ums, 7.7 \ums, and 11.3
\ums). These are thought to arise in photodissociation regions around
hot stars and are so common in starbursts that they are used as a
starburst vs AGN spectral classifier, as the PAHs are easily destroyed
by the harder radiation field from an AGN accretion disk
(e.g., \citet{genzel98,lut98}). Laurent et al. (2000) extend this
scheme to three templates (AGN, PDR and HII) to discriminate between a
pure AGN component, quiescent star forming regions with strong PAHs
associated with the PDR component and a starburst component with weak
PAH emission (i.e HII-like emission).  The mid-infrared (MIR)
continuum can also have a silicate absorption feature (rest wavelength
$\sim$ 9.7 $\mu$m) from an embedded AGN within an optically thick
torus, or from optically thick dust cocoons around hot stars. An
embedded source is also likely to have ice and gas absorption features
(e.g., Spoon et al. this issue). PAH and silicate features as
strong as those in many nearby galaxies would allow determination of
redshifts to z $\sim$ 3 for galaxies as faint as $\sim$ 1 mJy at 16
$\mu$m using the low resolution modules of the Infrared Spectrograph
on Spitzer (Houck et al. this issue). At z $\ga$ 1 these objects are
likely to be be ultraluminous infrared galaxies (ULIRG) with L$_{IR}
\sim 10^{12}$\ldot and hyper-luminous infrared galaxies (HLIRG) with
L$_{IR} \ga 10^{13}$\ldot. If these objects are similar to those found
in the local universe we may expect to find a dominent AGN in sources
with L$_{IR} \ga 10^{12.4}$\ldot \citep{tran01}.

It is anticipated that many important discoveries from Spitzer will
derive from wide area surveys, which will reveal large numbers of infrared
galaxies that are too faint for optical follow-up. Hence, the
capability to measure faint mid-infrared (MIR) spectra is crucial.

Observations of CFRS 14.1157 were taken as a first test of the IRS
long-wavelength low resolution spectrograph's (IRS-LL) ability to
measure the redshift of faint z $>$ 1 galaxies using MIR spectral
features and to characterise the MIR luminosity. The serendipitous
observation of CFRS 14.9025, a galaxy at z = 0.155 having flux only
0.35 mJy at 16 $\mu$m, further demonstrates the ability of the IRS to
observe faint sources.

CFRS 14.1157 is a galaxy identified with the radio source 15V23
(Fomalont et. al. 1991). The source is at a z of 1.15 (Hammer
et. al. 1995), which corresponds to a luminosity distance of 7900 Mpc
(H$_{o}$ = 71 kms$^{-1}$Mpc$^{-1}$, $\Omega_M$ = 0.27,
$\Omega_{\Lambda}$ = 0.73, $\Omega_{k}$ = 0) and an angular scale of 8
kpc per arcsecond. The source was detected by ISOCAM with a 15 \um
flux of 2.3 $\pm$ 0.08 mJy \citep{flo99}. However, the source appears
extended and revised photometry indicate that its flux is 3.1 $\pm$
0.3 mJy (Flores private communication). The HST I band image shows a
multi-component source embedded in diffuse emission and extended over
a region of $\sim$ 2.5 '' (20 kpc) \citep{webb03}.  \citet{was03}
detected x-ray emission with XMM-Newton and inferred the presence of a
fairly heavily obscured AGN from the 0.5-2/2-10 keV band hardness
ratio of $\sim$ -0.3. However, a comparison between the observed submm
flux and that predicted from the AGN led \citet{was03} to conclude
that there is a strong starburst component present.

The empirical limits which can be reached by the IRS in searching
for spectral features in faint sources and techniques for optimal
spectral extraction and template fitting using the SMART analysis
package (Higdon et al. 2004) are described in Section 2. The results
and the nature of CFRS 14.1157 are discussed in Section 3. The
conclusions are presented in Section 4.

\section{Observations and Analysis}

The IRS-LL module provides low resolution spectra (resolving power, 64
$\le$ $\frac{\lambda}{\Delta\lambda}$ $\le$ 128) between 14.0 $\mu$m
and 21.3 \um in long-low 2 (IRS-LL2) and 19.5 \um - 38.0 $\mu$m in
long-low 1 (IRS-LL1).  The observation was made on the 5$^{th}$ of
January 2004 in the IRS Staring Mode AOR, which integrated for 6
$\times$ 120 seconds at each of the two nominal nod positions on each
of the two slits (for observing mode details see chapter 7 of the
Spitzer Observers Manual
(SOM)\footnote{\url{http://ssc.spitzer.caltech.edu/documents/som/}}).
Therefore, the total integration time required to obtain the entire
spectrum shown in Figure 1 was 2880 seconds.

%FIGURE 1 CFRS 14.1157 SPECTRA
\begin{figure}
\epsscale{1.1}
\plotone{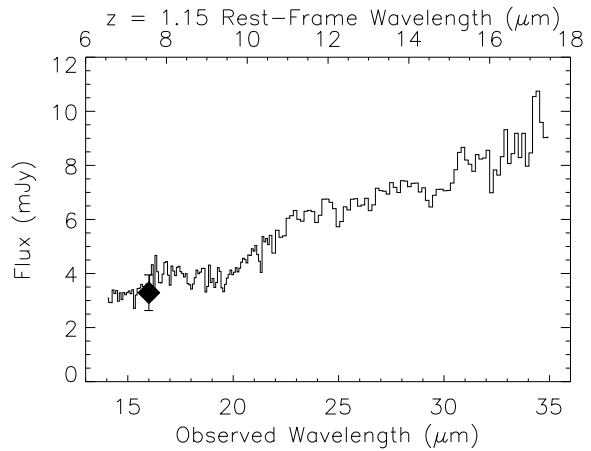}
\caption{ IRS-LL spectrum of CFRS 14.1157 (z = 1.15) with total integration
time of 2880 seconds. The diamond symbol is the blue-peakup flux.}
\end{figure}

In order to accurately place the source on the slit, CFRS 14.1157 was
first observed in the IRS blue peak-up camera. This also gives an
independent measure of the flux at 16 \ums. The peak up images
revealed two serendipitous detections of the galaxies CFRS 14.9025 and
CFRS 14.1129. The slit orientation fortuitously included CFRS 14.9025,
see Figure 2. The peakup photometry was based on the three double-correlated 
sampled images which were co-added on board and were obtained when the science
target was in the sweetspoot location of the camera (see the SOM).
CFRS 14.1157 has a blue peak-up 16 \um flux of 3.29 $\pm$ 0.66 mJy,
which agrees with the revised ISOCAM 15 \um flux. The source
positions and 16 \um fluxes are listed in Table 1. The spectral data
were processed as far as the un-flatfielded two dimensional image
using the standard IRS pipeline (see the SOM). The spectra were then
extracted and sky subtracted using the SMART analysis package (for
details see Higdon et al. 2004).

%FIGURE 2 PEAKUP 
\begin{figure}
\epsscale{0.8}
\plotone{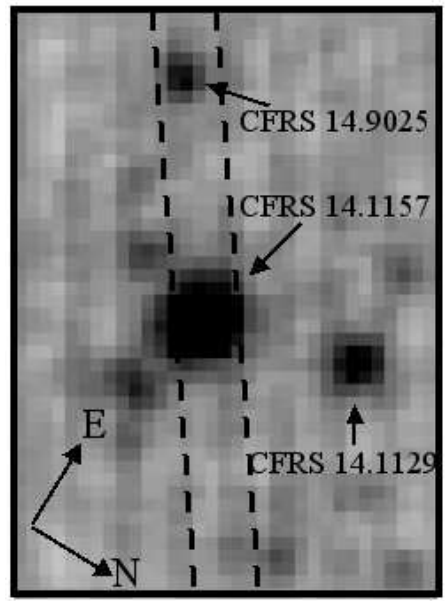}
\caption{Image from the Blue Peak up Camera on the IRS, having field
of view 60" by 72" and integration time 48 seconds.  Dashed line is
the LL slit position generated by the Spitzer Planning Observations
Tool (SPOT).}
\end{figure}

\begin{table}
\begin{center}
\caption{CFRS Sources in IRS Blue Peakup}
\begin{tabular}{lccccc}
\tableline

CFRS    &  RA\tablenotemark{a}& Dec\tablenotemark{a}&   z \tablenotemark{a} & F$_{\rm 16 \mu m}$\tablenotemark{b} \\
        &(J2000)&(J2000)&       &   (mJy)\\
14.1157 & 14:17:41.80 & 52:28:24 & 1.149 &3.29 $\pm$ 0.66\\
14.9025 & 14:17:44.99 & 52:28:03 & 0.155 &0.38  $\pm$ 0.08\\
14.1129 & 14:17:42.60 & 52:28:46 & 0.831 &0.66 $\pm$ 0.13\\
\tablenotetext{a} {\url{http://www.astro.utoronto.ca/~lilly/CFRS/}}
\tablenotetext{b} {IRS Blue Peak up flux.}
\end{tabular}
\end{center}
\end{table}

The extracted spectra were flat-fielded and flux-calibrated by
extracting and sky subtracting un-flatfielded observations of the
calibration star del UMi (from the same campaign) and dividing these
data by a del Umi template \citep{coh03} to generate a 1-dimensional
relative spectral response function (RSRF). The RSRF was then applied
to the observations of CFRS 14.1157 and CFRS 14.9025 to produce the
final spectra. The spectra of CFRS 14.1157 and CFRS 14.9025 are shown
in Figures 1 and 3, respectively. Their 16 \um fluxes are 3.6 mJy and
0.35 mJy, respectively, which matches the peak-up fluxes to 15 \%. The
integration time for CFRS 14.9025 is only half that for CFRS 14.1157
because the former object was present only in one of the two nod
positions. In the observation of CFRS 14.1157 the 16\um rms noise is
0.31 mJy and the 26 \um rms is 0.25 mJy. In CFRS 14.9025 the noise is
similar but the S/N is lower (see Table 2).

%FIGURE 3 CFRS 14.9025 SPECTRA
\begin{figure}
\epsscale{1.1}
\plotone{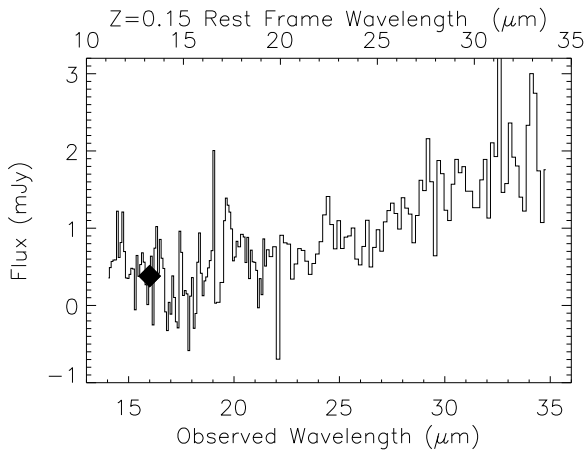}
\caption{IRS-LL spectrum of CFRS 14.9025 (z = 0.155) with total integration
time of 1440 seconds. The diamond symbol is the blue-peakup flux. }
\end{figure}

\begin{table}
\begin{center}
\caption{IRS Low Resolution Spectral Characteristics}
\begin{tabular}{lcccc}
\tableline

CFRS    & S/N$_{16 \mu m}$&RMS$_{16\mu m}$\tablenotemark{a} & S/N$_{26\mu m}$& RMS$_{26\mu m}$\tablenotemark{a}\\
&&(mJy)&&(mJy)\\
14.1157 & 12&0.31&26&0.25\\
14.9025 & 1&0.35&4&0.19\\
\tablenotetext{a} {RMS noise calculated from 0-order polynomial fit to the continuum between 15 and 17 \um and  25 and 27 \ums, respectively.}
\end{tabular}
\end{center}
\end{table}

The redshift was determined using a simple template fitting algorithm in
which infrared spectra of starburst galaxies or AGN were first
redshifted over the interval 0$<$z$<$3 in steps of 0.04, regridded,
scaled, and then compared to the IRS source spectra. Only discrete
emission and absorption features are fitted, hence the shape of the
continuum is treated as a free parameter. The value of z that
minimized the reduced $\chi^2$, defined as

\begin{equation}
{\rm
\chi^{2}_{\rm z} = N_{\rm free}^{-1} ~~\Sigma~\sigma^{-2}~(~D_{\lambda} -
a(~T_{\rm (1+z)\lambda}~+~b_{\lambda}~)^2
}
\end{equation}

was taken to be the source's redshift.  In this equation,
D$_{\lambda}$ and T$_{\rm (1+z)\lambda}$ are the input IRS source and
redshifted template spectra, respectively.  Scaling of the template
spectrum is managed by the constants a and b$_{\lambda}$.
The uncertainty in z is estimated by numerically calculating
                                     
\begin{equation}
{\rm
\sigma_{\rm z}^2 = \frac{2} {\partial^{2}\chi_{z}^{2}/\partial z^{2}}.
}
\end{equation}
                                                              
Because the CFRS spectra were obtained only in the second IRS
campaign, a sufficient number of sources have not yet been observed
with the IRS to produce a detailed catalog of starburst and AGN
templates. For this paper the templates are the ISO-SWS spectra
of M~82 and NGC 1068 \citep{sturm00} convolved to the IRS-LL
resolution.

\section{Results}

The CFRS 14.1157 MIR spectrum is featureless apart from a broad
absorption dip at $\sim$ 19 \um (see Figure 1). There is no measurable 
PAH emission. The spectrum of CFRS 14.9025 is also featureless, but
since the source is $\sim$ 8 times fainter than CFRS 14.1125 the 
signal to noise is low, meaning that if present, relatively strong
PAH emission features could be buried in the noise (see Table 3).

\begin{table}
\begin{center}
\caption{Relative 7.7 \um and 11.3 \um PAH Strengths}
\begin{tabular}{lcc}
\tableline &Relative Flux\tablenotemark{a}&RelativeFlux\tablenotemark{a}\\ 
&7.7 \um PAH&11.3 \um PAH\\
CFRS14.1157&$\le$0.3\tablenotemark{b}&$\le$0.1\tablenotemark{b}\\ 
M82&7.2\tablenotemark{c}&1.2\tablenotemark{c}\\
NGC1068&0.1\tablenotemark{c}&0.06\tablenotemark{c}\\

\tablenotetext{a} {PAH flux/continuum (11.6 - 11.9 $\mu m$). Note that the 
fluxes are in units of Wcm$^{-2}$.}
\tablenotetext{b} {The upper limit: 3-$\sigma \times$ observed bandwidth.}
\tablenotetext{c} {Sturm et. al. 2000.}
\end{tabular}
\end{center}
\end{table}

Using the spectral template method described in the previous section
and an AGN template, the minimal-$\chi_{z}^{2}$ redshift is 1.00 $\pm$
0.20 for CFRS 14.1157 (see Figure 4). The fit identifies the broad
absorption feature at $\lambda \sim$ 19 \um with silicate absorption
at $\sim$ 9.7 \ums, seen in NGC 1068. This is consistent with the
optically derived redshift of z = 1.15 $\pm$ 0.05 \citep{ham95}. In
Figure 5 the observed spectrum is shown overlayed with scaled versions
of the M82 and NGC 1068 templates redshifted to z = 1.0. The absence
of strong PAH features reveals that a typical starburst spectrum is
inconsistent with the observation of CFRS 14.1157. Moreover, the
continuum shape is clearly AGN like.

%FIGURE 4 chi2

\begin{figure}
\epsscale{1.1}
\plotone{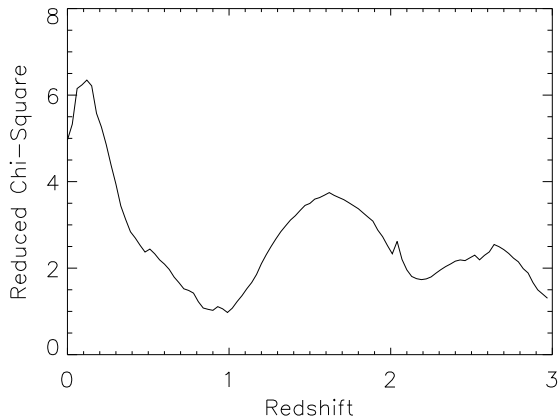}
\caption{Reduced $\chi^2$ for redshift determination of CFRS 14.1157 using NGC 1068 as a template.}
\end{figure}

\begin{figure}
\epsscale{1.1}
\plotone{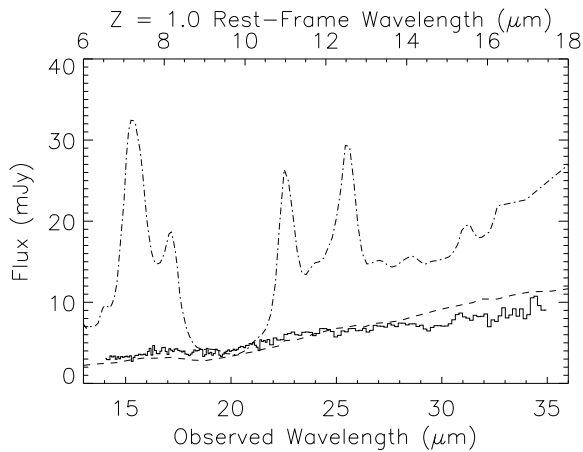}
\caption{CFRS 14.1157 overlaid with M82 (dot-dash line) and NGC 1068 
(dash line) templates at a redshift z = 1.0, determined by the MIR fit.}
\end{figure}

To estimate a formal limit to the relative 7.7 \um and 11.3 \um PAH
strength, the bandwidths for the PAHs are defined as 7.3 to 8.2 \um
and 11.1 to 11.7 \ums, respectively. A zero-order polynomial fit to
the continuum over the corresponding PAH band is fitted and the rms is
calculated. The integrated continuum between 11.6 and 11.9 \um is
measured. This is an ``off band'' region where no known spectral
features are found. The limit is then defined as three times the rms
multiplied by the PAH bandwidth and divided by the ``off band'' flux.
As shown in Table 3, 7.7 \um (11.3 \ums) PAH emission relative to the
continuum, must be at least 24 (12) times fainter than
observed in M82. In nearby galaxies the flux at 15 \um is strongly
correlated with the infrared luminosity, with L$_{IR} \sim 11.1$
L$_{15\mu m}$ \citep{cha01}. If this correlation holds for more
distant galaxies, then the IR luminosity of CFRS 14.1157 is $\sim$
10$^{13}$ \ldot, i.e. it is a hyper-luminous infrared galaxy.

The observed spectrum of CFRS14.9025 is featureless and its redshift
could not be determined using the IRS. At a luminosity distance of
731.4 Mpc the infrared luminosity, extrapolated from the 15 \um flux,
is $ 6 \times 10^{9}$ \ldot. Assuming that there is no AGN
contribution to the IR luminosity, this would correspond to a star
formation rate of $\sim$ 1 \mdot, using SFR [\mdot]
= 1.72 $\times 10^{-10}$ L$_{IR}$[\ldot] \citep{ken98}.

\section{Conclusions}

We have demonstrated the ability of the IRS-LL to determine the
redshift of a faint galaxy with f$_{16 \mu m}$ = 3.6 mJy. An rms of
0.3 mJy is reached with 2880 seconds of integration. The observations
indicate that CFRS 14.1157 is dominated by emission from an AGN
component in the MIR and belongs to the hyperluminous class of
infrared galaxies.

The serendipitous observation of CFRS 14.9025 demonstrates that we can
obtain a low resolution spectrum of a source at a flux level of 0.35 mJy
at 16 \ums. The IRS low resolution MIR capability will be crucial in
characterizing the nature and luminosity of newly discovered galaxies
at high redshift.

\acknowledgments This work is based [in part] on observations made
with the Spitzer Space Telescope, which is operated by the Jet
Propulsion Laboratory, California Institute of Technology under NASA
contract 1407. Support for this work was provided by NASA through
Contract Number 1257184 issued by JPL/Caltech.

%LIST ALL AUTHORS UNLESS MORE THAN 8

\clearpage

\end{document}